# Modeling of Multipath Transport

Chang Liu[1], Fei Song[2], Huan Yan[3], Sidong Zhang[4]

National Engineering Laboratory for Next Generation Internet Interconnection Devices
Beijing Jiaotong University, Beijing, China
[1] *deathsmile522@gmail.com*

[2] *fsong@bjtu.edu.cn*

[3] *10111005@bjtu.edu.cn*

[4] *sdzhang@bjtu.edu.cn*

**Abstract**
In this paper, we propose a model for evaluating the transmission performance of multipath transport. Previous researches focused exclusively on single pair users in simple scenarios. The distinct perspective in this paper is to build models for analyzing the performance when multipath transport is used in the entire network scope. We illustrate the influences on the transmission performance caused by the variation of network topologies, the services' arrival rate, the services' size and other parameters. We demonstrate through simulation that multipath transport could conditionally increase the throughput than single-path transport. And it has the capability to support higher services' arrival rate in various network topologies. And higher multi-parent probability will be beneficial for multipath transport to take its advantages.

## 1. Introduction

Multi-interface (3G, WLAN, WMAN, etc.) terminals which enable users to not only have access to services anywhere anytime from any network, but also through several interfaces, are increasing in user numbers, and can be expected to be the most common type of Internet device in the near future. Such trend has induced the emergence of the idea of 'multipath transport' which refers to the method of sending data over multiple available paths simultaneously. Multipath transport enables network resources to be used concurrently, and improves user experience. There are two key benefits of multipath transport. One is that the resilience of connectivity is strong with multiple paths, and the other is that it increases the efficiency of resource usage, and thus increases the network capacity available to end hosts [1]. As to the transport protocols at the transport layer, multipath functions are not new. Concurrent Multipath Transfer (CMT) [3-5] is a kind of multipath transport supported by the Stream Control Transmission Protocol (SCTP) [2]; Multipath TCP (MPTCP) [1,6] is a modified version of Transmission Control Protocol (TCP), whose original goal was to support multipath transport. Many subsequent researches on path selection [7-10], load sharing [11-14], retransmission judgment [15-16], throughput estimation [17-20], receiver buffer size [5,21], have been conducted to improve transmission performance of these two protocols.

However, the majority of the current works on multipath transport focus exclusively on the performance of single pair users, leaving out the consideration for the entire network topology and the possibility of multiple pairs of users. The simulations in these works were often set up based on environments con two terminals, and the two terminals accessed network though multipath. They paid more attentions on the performances between the single pair users, which is not close to the actual scenarios.

The motivation of this paper is to discuss the performance of multipath transport when it is used in the entire network by multiple pairs of users instead of just single pair. An analytic model was proposed in order to achieve this object. And we also analyze the multipath transport performances in different kinds of network topology based on analytic model.

In this paper, we pay more attentions to the performances of the entire network. Different from other researches, we analyze what will happen in the network if all of the endpoints use multipath transport, and make a comparison with all of them use single-path transport. First of all, we have made a topological model to construct the network topology, and made a services model to simulate the arrival services. With these two models, we have demonstrated that multipath transport could obtain higher throughput than single-path transport, especially when multipath transport is used in the entire network. And we have also demonstrated that multipath transport could support higher services' arrival rate than single-path transport. Secondly, with changing the network topology of simulations, we have demonstrated that higher multi-parent probability will be beneficial for multipath transport to take its advantages.

## 2. Multipath Transport and Single-path Transport

2.1 Single-Path Transport

At transport layer, TCP is a common protocol which provides reliable, ordered delivery. Due to the basic design principle, the original TCP does not support multihoming terminals, so it only supports single-path transport. Major Internet applications such as the World Wide Web, email, remote administration and file transfer rely on TCP, and single-path transport is widely used in current Internet. It can be said that the current Internet is based on this kind of single-path transport applications.

SCTP is another transport protocol which provides reliable, ordered delivery just as TCP. One key different between SCTP and TCP is SCTP could support multihoming terminals. This feature makes SCTP suits for modern network better than TCP. However, although SCTP supports multihoming terminals, the original SCTP still support single-path transport only. It could build more than one path among two terminals, but it only allows using one path to transmit data, and the other paths are backup paths. So in essence, the original SCTP is still using single-path transport.

2.2 Multipath Transport

Recently, the user requirements keep increasing rapidly, especially reflected in bandwidth. Single-path transport is hard to satisfy these situations, and it needs several single paths to combine together to provide better services. So multipath transport is presented.

Multipath TCP (MPTCP) [1] is a modified version of TCP. It adopts the idea of subflows to realize multipath transport, different subflows are sent to different paths. The mechanism of each subflow is just likes original TCP's. This research direction has been accepted by IETF [1].

Concurrent Multipath Transfer (CMT) [3-5] is based on SCTP. With the feature of supporting multihoming in SCTP, CMT builds multipath among single pair, and uses these path to transmit data concurrently, thus it realizes multipath transport.

Both of them are hot topics. There are many researches related with multipath transport (MPTCP and CMT), such as the problem of out of order packets, lost packet and retransmission judgments, receiver buffer size, load balance among multipath, etc.. These researches have already demonstrated that multipath transport can provide better end to end services than single-path transport could.

However, multipath transport means it will seize more network resources. Although multipath transport can improve end to end services, if each user uses multipath transport, it is difficult to predict the transmission effect. So before multipath transport is used widely in the Internet, it needs to do many analyses of the entire network with multiple pair multipath transport users.

2.3 Load Sharing in Multipath Transport

The load sharing is one of the main questions in multipath transport. In multipath transport, there are several available paths we could use, how to use them, how many data will assign to them is a real question. So many researches are about this question, but they have different emphases, such as path selection [7-9] (it is about how each packet selects its sending path), data distribution [10-11], load balance [12-14], etc.. Most of those researches have a consensus, the wider path should send more data, and this also coincidences the common understanding.

Therefore, in our paper we assume that multipath transport assigns data to each path based on the ratio of real-time bandwidths.

## 3. Construction of Network Topology

3.1 Node Classification

Here, we propose a classification method for the nodes in the current Internet structure. We divide the nodes into different levels based on each's switching performance. For instance, a 10Gb/s node can be classified into core level, and a 10Mb/s node can be classified into leaf level. Nodes at different levels also have different topological properties. Higher level nodes often have higher Connectivity Degree, which means that they have more connections.

For convenience of demonstration, we create a three-level network structure. Note that in reality, the network structure created using our method is not limited to three levels. In the structure, the nodes at the lowest level are leaf nodes, while the rest are core nodes.

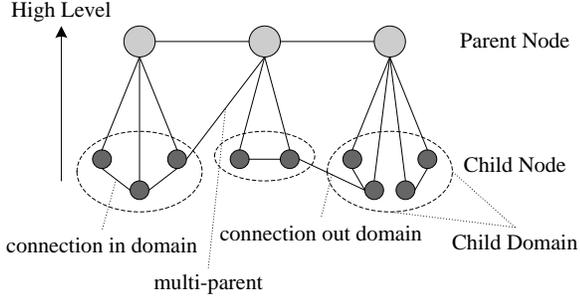

Fig.1 Different level nodes in networks

Nodes are also classified as parent/child in our structure. The definition and characteristics of the parent/child nodes are as follows:

- A parent node connects to its child nodes, and operates at the level atop them. A child node that is not at the bottom level of the network structure can also be the parent of its own children.
- The number of child nodes connected to a parent node is random, and we assume this number obeys to the uniform distribution. $\mu_l^{NC}$ is used to express the mean value of the number at the level $l$. This parameter is majorly determined by the scale of the network.

The relationship between parent nodes and child nodes is shown in Fig.1.

### 3.2 Domain

A domain is defined as the set of all the child nodes connected to a parent node. One example of a domain on the Internet would be a subnet. The set of the terminals that are using the same network interface can also be considered as a domain. For instance, the computers using wired access mode are in a different domain from the ones using wireless access mode, although they might be close in distance. In this paper, the probability of a connection between any two nodes depends on the domains they are in. The features of a domain are as following:

- The value of the probability of a connection between any two nodes within the same domain is based on their network level. $P_l^{in}$ is used to express this probability at the level $l$.
- Similarly, $P_l^{out}$ is used to express the probability of a connection between any two nodes that are at the same level $l$, but in different domains.
- In our network structure, the more distance there is between any two nodes in different domains, the smaller chance there is that they will directly connect. We propose a threshold value $D_{thrsh}$, so that the probability of a connection between them $P_l^{out}$ will become invalid when the distance between any two nodes falls below this value.
- In general $P_l^{in}$ of any node is greater than its $P_l^{out}$. These two parameters are directly influenced by the topology of the network. For instance, a mesh network will see a higher probability of a connection between the nodes than a star network or a tree network.

### 3.3 Multi-parent

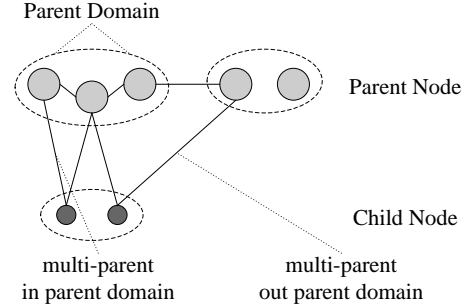

Fig.2 Multi-parent probability

It is also possible that a node connects with multi-parent nodes, especially when it is at high level. In the past, a terminal usually had only one network interface, or was only able to access the network through one interface at a time, and thus the number of multi-parent nodes was small. However, as multi-interface terminals have become widely used in the past decade, multi-parent nodes are now expected to become the major network nodes in the near future.

Here, we use $P_l^{MP}$ to express the overall probability of a node at the level $l$ connecting to multiple parent nodes at the level $l-1$.

Still, it is difficult for a node to connect with multiple parents that are too distant from each other at the same time. Therefore, we will only consider two types of multi-parent connection, as shown in Fig.2:

- For a child node, its first parent node is defined as its original parent node, and it is set during the creation of the node. Then we define the domain of its original parent node to be its parent domain. Any node within this domain will have a probability $P_l^{MPin}$ to be its parent node.
- A child node can only connect with a new parent outside of its parent domain, when the outside parent to be is directly connected with the original parent node. When this condition is fulfilled, this probability is valid and is expressed as $P_l^{MPout}$.

a) Make random number of Level 1 nodes, average number = $\mu_1^{NC}$;

b) for each Level 1 node:
    b.1) make random number of Level 2 nodes and connect with the parent node, average number = $\mu_2^{NC}$;
    b.2) for each Level 2 node:
        b.2.1) has a probability of $P_2^{MPin}$ to connect with another Level 1 parent node;
    b.3) in the domain, two Level 2 nodes have a probability of $P_2^{in}$ to be connected;

c) outside the domain, two Level 2 nodes have a probability of $P_2^{out}$ to be connected;

d) for each Level 2 node:
    d.1) make random number of Level 3 nodes and connect with the parent node,, average number = $\mu_3^{NC}$;
    d.2) for each Level 3 node:
        d.2.1) has a probability of $P_3^{MPin}$ to connect with another Level 2 parent node which is in its parental domain;
        d.2.2) has a probability of $P_3^{MPout}$ to connect with another Level 2 node which is outside its parental domain;
    d.3) in the domain, two Level 3 nodes have a probability of $P_3^{in}$ to be connected;

e) outside the domain, two Level 3 nodes have a probability of $P_3^{out}$ to be connected.

Fig.3 Pseudocodes for network topology construction

### 3.4 Parameters of the Network Structure

Summing up the above, there are five major parameters in our network structure: $\mu_l^{NC}$, $P_l^{in}$, $P_l^{out}$, $P_l^{MPin}$ and $P_l^{MPout}$. By varying the parameters, we could obtain and run simulations for different network topologies.

### 3.5 Process of Constructing Network Topology

Basing on the above analyses, we can now construct network topologies using the steps shown as pseudocodes in Fig.3.

## 4. Modeling for Service Transmission

Internet network services occupy network resources and create data flows from one terminal to another through a series of routers. In this paper, we built a model for such data flows in order to analyze their transmission process.

We assume that a service is provided by a leaf node, transmitted through a series of nodes, and in the end received by another leaf node. This assumption is based on the fact that in our model, higher level nodes are the abstractions of the real-world routers, and the leaf nodes represent real-world terminals. We refer to the routers (higher level nodes) a particular data flow has travelled through as a path. According to the common routing rules, each node has a specific transmission capability, which reflects the node's maximum forwarding rate. It's not hard to infer that a path's bandwidth is the minimum of all the nodes' bandwidth along this path. All the data flow passing through a node will occupy its transmission capability averagely. We define a term 'services' size', which stands for the total size of the data packets a service creates. If a service has finished transmitting the data, it will release the network resources. Also, a random number of new services will arrive at a certain rate.

Based on the above model, simulations could be carried out both for single-path transport and multipath transport scenarios to examine their transmission performance and for further comparison.

### 4.1 Network Services

In our model, a number of new services will emerge on a fixed time basis. According to the classical queue model, we assume that the services' arrival rate obeys Poisson distribution P(λ), as shown in Formula (1). Adjustments to the services' arrival rate can be achieved by varying the parameter λ.

$$\text{Number of new services} \sim P(\lambda) \qquad (1)$$

### 4.2 Service Paths

In our model, we will adapt existing routing algorithms, such as OSPF, RIP, etc which will select a path with the minimum hop count, for both single-path and multipath transport.

In Fig.4, the arrows indicate the direction of the data flow. And the single path is the lines with the single arrow, which indicates the minimum hop count from the source to the destination. For the purpose of demonstration, we adopted a two-path transmission model. Since multipath transport is made possible by terminals accessing the network through different interfaces, the first and the last hops in our model should be separate. However, it is possible that the two paths cross each other at a certain node inside the network.

In our model, multipath transport finds its two paths by following steps:

- Find the path with the minimum hop count from the source to the destination, and mark it as Path 1;
- Mask the first and the last node in Path 1 temporarily;
- In the new topology, repeat Step 1, then mark the resulting path as Path 2, which will have different first and last hop from Path 1;
- If Path 2 cannot be found, multipath transport will not be carried out;
- Unmask the first and the last node in Path 1.

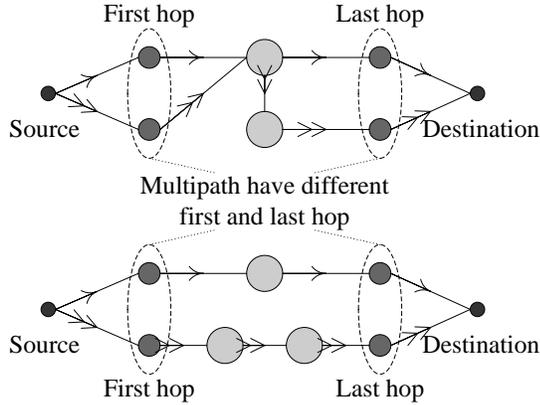

Fig.4 Two examples of multipath transport

Fig.4 shows the resulting two paths. One is marked by a series of single arrows, and the other by double arrows. Note that these two paths have separate first hops, and separate last hops, too. In each path, there will be an independent flow for the specified service.

### 4.3 Transmission Capability of The Nodes

In actual networks, routers receive packets and forward them based on its routing algorithms, and their forwarding rate can suffice as an abstracted representation of a router in a mathematical model because it is usually considered to be the most fundamental property. In our particular model, this abstracted property will be called transmission capability of a node, and nodes at a higher level will have greater transmission capability due to our node classification method. Also, transmission capability will be the same for the node that are at the same level.

According to 4.2, there will be multiple data flows coming from different services passing through one node. We assume all services are equal users of a node's transmission capability splitting its resources evenly. For this reason, we introduce the concept of a node's bandwidth, which stands for the value of transmission capability a service occupies at a certain node. Its mathematical manifestation is shown in Formula (2).

$$b_i^{node} = \frac{C_i^{node}}{Num_i^{node\_flow}} \quad (2)$$

In formula (2), $i$ is the serial number of a certain node; $b_i^{node}$ is the node's bandwidth of Node $i$; $C_i^{node}$ is the transmission capability of Node $i$, and $Num_i^{node\_flow}$ is the number of data flows passing through Node $i$.

### 4.4 Transmission Throughput of A Service

The transmission throughput of a service's data will be equal to the lowest node bandwidth along the transmission path, as Formula (3) shows. $b_p^{path}$ is the bandwidth of Path $p$, and $b_i^{node}$ is the bandwidth of Node $i$.

$$b_p^{path} = \min_{i \in Path_p}(b_i^{node}) \quad (3)$$

It is not difficult to see that, for single-path transport, the transmission throughput of a service's data is equal to the value of its path's bandwidth.

For the multipath transport in our model, we will discard the interaction between the two paths because they can be reduced using many available methods, such as using optimized transport protocol, enhancing the receiver buffer's capacity, etc. Therefore, in our model, the transmission throughput of a service's data with multiples paths can be expressed as the sum of all paths' bandwidth, as Formula (4) shows.

$$TP = \sum_p b_p^{path} \quad (4)$$

### 4.5 Size of A Service

The size of a service here refers to the overall size of all the data the service needs to transmit. In our model, we assume that this property obeys Uniform Distribution, as Formula (5) shows. $S_k^{t_0}$ is the initial size of a certain service $k$; $\mu_S$ is the mean value of all the data sizes a service could produce.

$$S_k^{t_0} \sim U(0.5\mu_S, 1.5\mu_S) \quad (5)$$

We have also considered alternatives of data size distribution pattern, e.g. Exponential Distribution and Constant Value Distribution when running the simulations, but the results were very similar. Therefore, we only demonstrate through the uniform distribution in this paper.

When all the data of a certain service has reached its destination, this service will terminate itself and release the path.

## 4.6 Final Model for Service Transmission

In each time unit, a service will transmit a bulk of data, whose size is equal to the value of the transmission throughput of this service from the origin to the destination, and the size of the residual data will decrease correspondingly until it has reached 0, meaning this service is done and is closing itself.

We will use $S_k$ to express the size of the residual data of Service $k$, and thus the iterative formula could be expressed as Formula (6). $S_k^{t+1}$ and $S_k^t$ are the sizes of the residual data at the time $t+1$ and the time $t$; $TP_k^t$ is the transmission throughput at the time $t$.

$$S_k^{t+1} = S_k^t - TP_k^t \qquad (6)$$

Finally, with the formulas from this chapter, we were able to build a mathematical model for network service transmission.

$$\begin{cases} S_k^{t_0} \sim U(0.5\mu_S, 1.5\mu_S) \\ S_k^{t+1} = S_k^t - \sum_j \min_{m \in \text{Path}_j} (\frac{C_m^t}{\text{node\_flow}}) \\ \phantom{S_k^{t+1} = S_k^t - \sum_j \min_{m \in \text{Path}_j}} Num_m^t \end{cases} \qquad (7)$$

## 4.7 Service Modeling Processes

We summarized the processes for building this model, as Fig.5 shows.

For each unit time:

a) create random number of new services which obeys $P(\lambda)$;

b) for each new service:
    b.1) source and destination are choosing from Level 3 nodes randomly;
    b.2) find 1 or 2 shortest path for the service; (for single-path transport or multipath transport)
    b.3) add the number of flows for each node which is in the shortest path;

c) for each service:
    c.1) calculate the bandwidth of its path;
    c.2) calculate the throughput of service at this unit time;
    c.3) decrease service size due to its throughput;
    c.4) if service size $\leq 0$, delete the service and the flows in its paths;

d) end one unit time.

Fig.5 Pseudocodes for Service Transmission Modeling

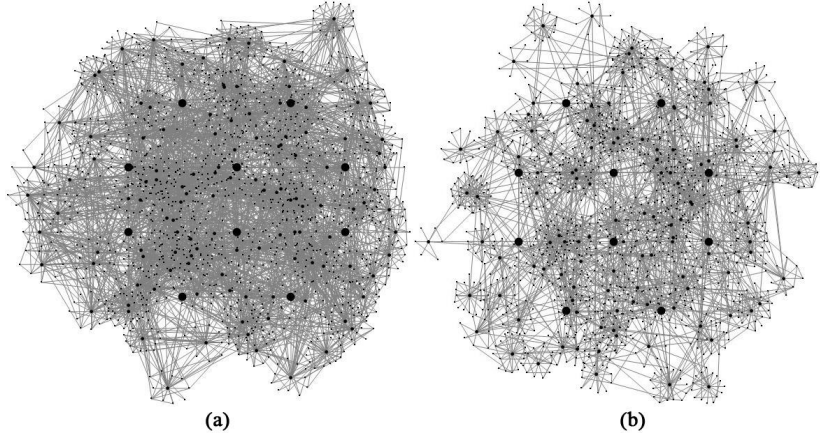

(a)        (b)

Fig.6 Network Topologies for Simulations

Table 1 Parameters of the Network Topologies for Simulations

| Level of the Nodes | Number of Child Nodes | Probability of A Connection | | Probability of Multi-parent Scenarios | |
|---|---|---|---|---|---|
| | $\mu_l^{NC}$ | $P_l^{in}$ | $P_l^{out}$ | $P_l^{MPin}$ | $P_l^{MPout}$ |
| Level 1 | 10 | 100% | - | - | - |
| Level 2 | 10 | 20% | 5% (range 30) | 5% | - |
| Level 3 | 10 | 10% | 2% (range 30) | 10% | 5% |

# 5. Simulations and Results

This chapter discusses the network simulations we ran and the corresponding results we obtained. 5.1 talks about the construction of network topologies in the simulation environment according to the node classification methods in Chapter 3. 5.2 compares the single-path transport and the multipath transport in many aspects under the topologies we constructed. 5.3 discusses the effects of varying network topology on the service's models.

5.1 Constructing the Network Topology

According to Chapter 3, we constructed the network topologies for simulation network with 3 levels of nodes, where Level 1 is the highest and consists only of the core nodes, and Level 3 is the lowest and consists only of the leaf nodes. The parameters for constructing these network topologies were set according to Table 1.

Fig.6(b) demonstrates a topology with a low multi-parent probability, which was obtained by multiplying the multi-parent probability of all the nodes by a factor of 0.2. This topology is very similar to the actual network topologies of today, because, though having multiple network interfaces, current terminal devices usually solely use one interface to access the networks. On the other hand, Fig.6(a), which was created using larger multi-parent probability, will better model a topology where terminals accessing the network through multiple interfaces simultaneously.

5.2 Simulation for Single-path Transport and Multipath Transport

In the simulations, the unit of time was set as 1 time unit, and the unit of services' size as 1 size unit, and therefore, the transmission throughput is 1 size unit per time unit. All simulation time length is 1000 time units, and all results are average values of the properties tested in these 1000 unit times. Using Mathematic 8.0, we ran simulations of SP (Single-path Transport) and MP (Multipath Transport) for the topologies we built.

A) Service Transmission Throughput

In our simulations, all services were set to have certain throughputs (Part 4.4), and we could get the average value of service throughputs for each unit time, which is determined by the services' arrival rate $\lambda$ (Part 4.1).

Fig.7 shows the service throughput at each time step, under $\lambda=300$ and $\lambda=340$, respectively. We concluded from the results that:

- Under $\lambda=300$ (Fig.7(a)), the network is steady. The service's throughput varies but is bounded. This phenomenon distinguishes from that under $\lambda=340$ (Fig.7(b)), where the SP transport failed to keep the network state steady, which is demonstrated through a graduate decline in the services' throughput. The mechanism for the decline is that a $\lambda$ as high as 340 is 'unbearable' to this network topology with SP transport. The services' arrival rate becomes higher than the services' finishing rate, resulting in a traffic situation. However, with MP transport, even under a $\lambda$ as high as 340, the throughput managed to stay steady.
- It can be observed that the services' throughput keeps decreasing until reaching the steady range. This is because the network is set to start with no services at all, and it takes a certain period of time for it to go from its initial state to the steady state. This transient state takes about 20~50 time units, and depends on a number of factors such as the service's size. Further analyses of the factors will be made later in this paper.
- The diagrams show that in this particular network topology, MP transport yields higher service throughput than SP transport. Detailed analyses will be made later in this paper.
- It can also be inferred that, with the same topology and the same $\lambda$, MP transport tend to accommodate higher services' arrival rate than SP transport. Detailed analyses will be made in later chapters.

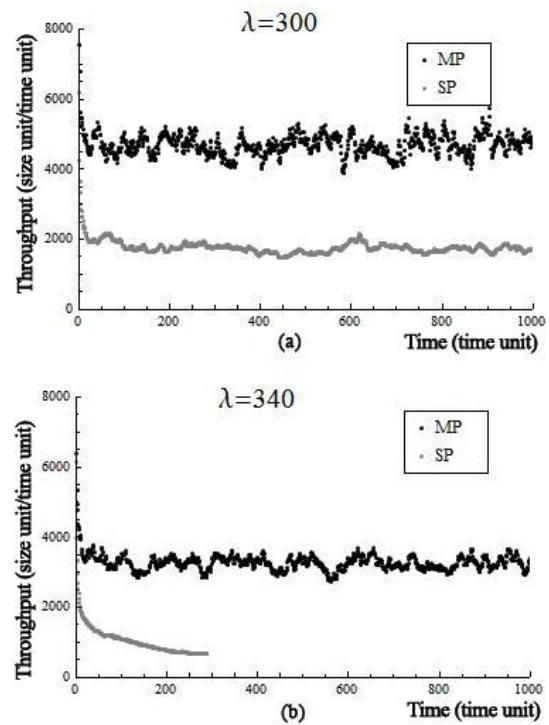

Fig.7 Real-time services' throughputs

In Fig.8, we demonstrate the relationship between the services' throughput and the services' arrival rate. It can be observed that the MP throughput is almost twice as large as the SP throughput at the same arrival rate in our specific topology. We believe that this can further support our inference that in the same network topology and with the same λ, using MP transport tends to result in higher service throughput than using SP transport.

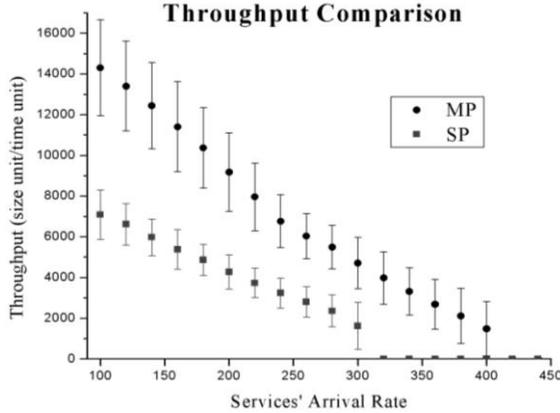

Fig.8 The Services' Throughput With the Services' Arrival Rates

### B) Services' Size

Fig.9 demonstrates the influence of the services' size $\mu_s$ on the services' arrival rate in our model. It can be inferred from Fig.9 that MP tends to support higher services' arrival rate than SP. Further analysis reveals that the value of the product $\lambda_{max} \bullet \mu_s$ remains constant. For MP, $\lambda_{max} \bullet \mu_s \approx 4150$, whereas for SP, $\lambda_{max} \bullet \mu_s \approx 3050$. This constant in fact reflects the overall transmission capacity of a network topology under a certain type of transport method. In our network topology, MP increases this capacity by about 35% compared to that of the SP. Note that in later chapters of this paper, the simulations were done with the services' size 10.

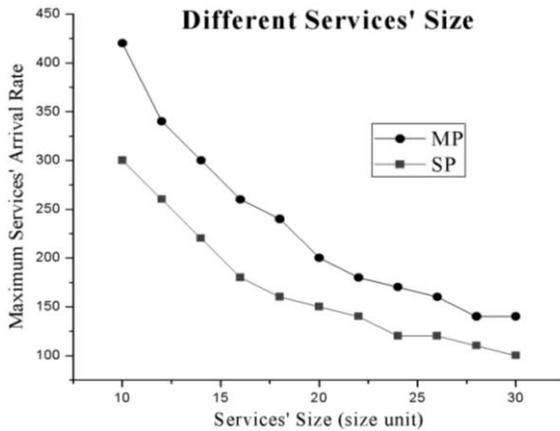

Fig.9 Different Services' Size

### C) MP Transport In Single Pair and Entire Network

As Chapter 1 stated, most previous researches focused solely on testing their transport strategies through simulations that were set in a network with only a single pair users, which is not representative of actual networks. Such researches were often conducted in a multivariate analysis fashion, where the network topologies and the service models were held constant, with the transport strategy as the only variable, in order to demonstrate the resulting superior network performance. Here, we adapted a similar fashion in which we experimented with our MP transport strategy, but what is different is that here, it is both the network topology and the transport strategy that are the variables. We applied both the MP and the SP transport strategies respectively to a simple single pair ends, and later to the entire network. The simulation results we obtained are shown in Fig.10.

Fig.10 shows the Throughput Gain (TG), which refers to the ratio of MP throughput to the SP throughput when applied to the same network topology, with respect to the services' arrival rate. This parameter measures the improvements the MP brought compared to the SP. It can be observed that, When SP and MP was applied to a single pair users, the TG reached as high as 1.8, but still, it fails to grow over 2.0. This is due to the fact that our MP strategy, which introduces only one additional path, can increase the overall throughput by 100% at maximum. The fact that the TG was always above 1.0 is enough to demonstrate that the MP is superior to SP in a simple two-end network. This result stays true for the more complex network as well. However, When SP and MP was applied to the entire network, it can be observed that as the services' arrival rate grows, the TG grows over 2.0, and at one point reached as high as 3.0. This is because when MP is applied, the data flow within the entire network can be balanced, enhancing the bandwidth along each path.

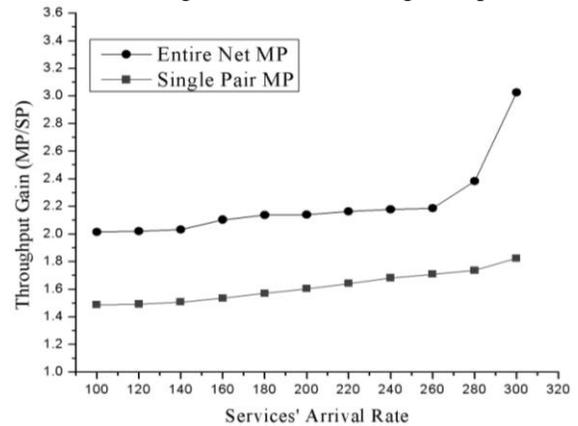

Fig.10 TG in Single Pair Users and Entire Network

## 5.3 Effect of Variations in the Parameters of the Network Topologies

### A) Effects of Variation in the Multi-parent Probabilities

Table 2 Multi-parent Probabilities Variation Range

| Level | $P_l^{MPin}$ | $P_l^{MPout}$ |
|---|---|---|
| Level 2 | 4%~20% | - |
| Level 3 | 4%~20% ($P_2^{MPin}$) | 2%~10% ($P_2^{MPin}/2$) |

Multi-parent probabilities $P_l^{MPin}$ and $P_l^{MPout}$ were introduced in 3.3. A low multi-parent probability indicates that an endpoint is less likely to use multiple interfaces to access the network, so it will be hard to find the second path between the source and destination. The aim of this part is to measure the impact of multi-parent probabilities. The variation range we set for the multi-parent probabilities is shown in Table 2.

Fig.11 shows the effects of multi-parent probability $P_l^{MPin}$. From our observation, it can be inferred that the higher the multi-parent probabilities grow, the higher the TG gets, because when the value of $P_l^{MPin}$ is small, the superiority of MP are seriously compromised as the endpoints find it hard to connect through multiple paths. However, the TG reached a plateau (about 3.276) when $P_2^{MPin}$ hits 10%.

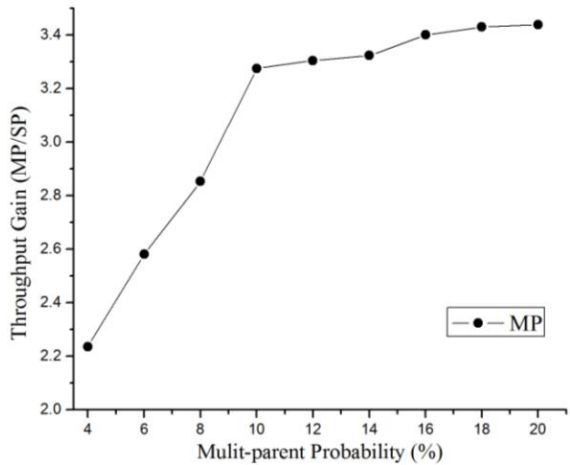

Fig.11 TG With Respect To $P_2^{MPin}$

### B) Effects of Variation in the Number of the Child Nodes

The quantity of the child nodes $\mu_l^{NC}$ directly influences the scale of the network and the loading pressure of higher level nodes, as 3.1 stated. Here, we aimed to examine the effects of $\mu_l^{NC}$. The variation range of $\mu_l^{NC}$ is set to be 6~16 (under the basic network topology, with $\mu_l^{NC}=10$).

It can be observed in Fig.12 that the larger the child quantity gets, the higher the services' arrival rate for both MP and SP transport there is, bringing a higher transmission capability. The MP, however, is still superior to SP at all times. The relationship between the child quantity and the maximum services' arrival rate is close to being linear.

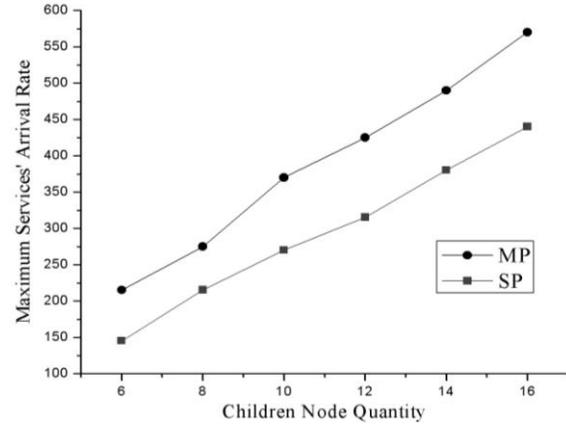

Fig.12 Maximum Services' Arrival Rate With Respect to the Child Node Quantity

## 6. Conclusions

In this paper, we proposed a topological model to construct the network topology, and created a services model to simulate arrival services. The simulation results demonstrate that, 1) MP could obtain higher throughput than SP. Moreover, if MP is used in the entire network, the improvement of throughput will be more remarkable, for the throughput gain of entire net MP is about 2.1, but the gain of single-pair MP is only about 1.5. 2) MP could also support higher services' arrival rate than SP, and the improvement is about 30%. 3) Services' size will affect the maximum services' arrival rate, but $\lambda_{max} \bullet \mu_s$ will remain a constant value which reflects the overall transmission capacity of a network topology under a certain type of transport method, and MP could increase this capacity by about 35% compared to SP. 4) When multi-parent probability is lower than 10%, there is a linear relationship between multi-parent probability and throughput gain, so higher multi-parent probability will be beneficial for MP to take its advantages. 5) Larger network size will support higher services' arrival rate, and there is also a linear relationship between network size and maximum services' arrival rate, for both MP and SP.

**Acknowledgments**

This paper is supported in part by the National High-Tech Research and Development Program of China (863) under Contract No. 2011AA01A101, in part by the National Natural Science Foundation of China (NSFC) under Contract No. 61232017, No. 61102049 and No. 61100217, in part by Ph.D. Programs Foundation of Ministry of Education of China under Contract No. 20120009120005.

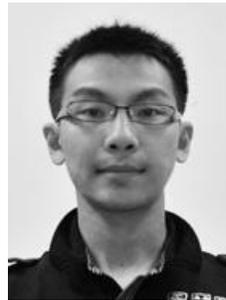

**Chang Liu** is a full-time Ph.D. candidate in School of Electronic and Information Engineering, National Engineering Laboratory for Next Generation Internet Interconnection Devices, Beijing Jiaotong University. His current research interests are Next Generation Internet Architecture, Transport Protocol and Network modeling.

**Fei Song**, received his Ph.D. degree from Beijing Jiaotong University in 2010. He is now a Lecturer in School of Electronic and Information Engineering, National Engineering Laboratory for Next Generation Internet Interconnection Devices, Beijing Jiaotong University. His current research interests are Next Generation Internet Architecture, Wireless Communications, Cloud computing.

**Zhang Sidong**, is now a professor in Beijing Jiaotong University. He has published more than 100 research papers in the areas of wireless communications, computer networks, Ad-hoc networks, sensor networks and information theory. Professor Zhang is also a member of the electronics and information science steering committee of the Ministry of Education, a member of the expert committee of the national Natural Science Foundation of China (NSFC).